\documentclass[pra,twocolumn,showpacs,superscriptaddress,a4paper]{revtex4}
\usepackage{graphicx}
\usepackage{psfrag}
\usepackage{amsmath}
\usepackage{amssymb}
\usepackage{amsfonts}

\newcommand{\vb}[1]{{\mbox{\boldmath$#1$}}}

\begin{document}
\title{States without linear counterpart in Bose-Einstein condensates}
\author{Roberto D'Agosta}
\email{dagosta@fis.uniroma3.it}
\affiliation{ Dipartimento di Fisica ``E.Amaldi'', Universit\`a di
Roma 3, via della Vasca Navale 84, Roma 00146 Italy}
\affiliation{Istituto Nazionale per la Fisica della Materia, Unit\`a 
di Roma III}
\author{Carlo Presilla}
\email{carlo.presilla@roma1.infn.it}
\affiliation{Dipartimento di Fisica, Universit\`a di Roma ``La Sapienza'',
Piazzale A. Moro 2, Roma 00185, Italy}
\affiliation{Istituto Nazionale per la Fisica della Materia, 
Unit\`a di Roma I} 

\date{\today}
\begin{abstract}
We show the existence of stationary solutions of a 1-D Gross-Pitaevskii 
equation in presence of a multi-well external potential that
do not reduce to any of the eigenfunctions of the associated 
Schr\"odinger problem. 
These solutions, which in the limit of strong nonlinearity have the 
form of chains of dark or bright solitons located near the extrema of 
the potential, represent 
macroscopically excited states of a Bose-Einstein condensate
and are in principle experimentally observable. 
\end{abstract}
\maketitle    

\section{Introduction}
Bose-Einstein condensation (BEC) of weakly interacting atomic gases 
\cite{isw} strongly motivates the study of the Gross-Pitaevskii 
equation (GPE), 
\begin{equation}
\left[-i\hbar\frac{\partial}{\partial t}
-\frac{\hbar^2}{2m}\nabla^2+U_0|\Psi(\vb{x},t)|^2+V(\vb{x})
\right]\Psi(\vb{x},t)=0,
\label{gpe}
\end{equation}
a mean-field Schr\"odinger equation with 
local cubic non-linearity.
Of particular interest are the non ground-state stationary solutions of 
the GPE \cite{yyb,kat} which represent macroscopically excited states 
of the condensate.
Vortices have been recently observed in two- \cite{matthews}
or one-component \cite{madison} condensates and are also invoked as a
superfluidity breaking mechanism \cite{onofrio}.
Phase engineering optical techniques have allowed to generate dark 
solitons in atomic gases with positive scattering length 
\cite{burger,denschlag}.

Vortices and solitons observed in recent experiments are examples of
excited states with linear counterpart, i.e. stationary solutions
of the GPE which can be obtained as a deformation of eigenstates of the 
corresponding linear Schr\"odinger equation \cite{dmp,ccr}. 
However, the GPE may also admit stationary solutions without linear 
counterpart.
In a discretized version of the GPE, also known as discrete self trapping
equation, the existence and stability of solutions without linear
counterpart has been studied at various discretization orders \cite{els}.
In particular, the appearance of self trapping stationary states 
in the dimer case, which mimics a double-well system,
has been widely investigated in connection with the 
evolution of wave packets \cite{sfgs,mcww,oklhca}. 
Recently, a set of stationary solutions without linear counterpart 
has been discovered also in the continuous case, 
namely the exactly solvable 1-D GPE with periodic boundary conditions 
and zero external potential \cite{ccr}.
These states break the rotational invariance of the associated linear
problem.

In this paper, we show the existence of stationary solutions 
without linear counterpart of a 1-D GPE in presence of a  
multi-well external potential. 
In the limit of strong non linearity, these solutions assume the form 
of chains of dark or bright solitons located near the extrema of the 
potential and in general break the symmetry of the external potential.

Our analysis is of direct interest for BEC experiments where 
atomic gases can be confined in arbitrarily tailored 
magnetic or optic traps. 
As a case study, we investigate a GPE representing a quasi 1-D 
Bose-Einstein condensate confined in a double-well trap
described by the potential
\begin{equation}
V(x)=m^2 \gamma^4 x^4-m\omega^2 x^2+\frac{\omega^4}{4 \gamma^4}.
\label{bistablep}
\end{equation}  
In Section \ref{analytical} 
we describe all the zero-, one-, and two-soliton 
solutions of this model in an analytical way valid in the limit 
of strong non linearity. 
In Section \ref{numerical} by means of numerical simulations 
we find the exact shape of these states and study their evolution 
in the linear limit reached when the number of 
particles in the condensate, $N$, vanishes.
We consider both the cases of condensates with positive or negative 
scattering length. 
Shape and energy of the corresponding stationary solutions are 
shown in Fig.s 1, 3 and 2, 4, respectively.
Their stability properties are discussed in Section \ref{stability}.
\begin{figure*}  
\psfrag{N=0}[r][r][0.9]{$N=0$}
\psfrag{N=0.15}[r][r][0.9]{$0.15$}
\psfrag{N=5000}[r][r][0.9]{$5 \times 10^3$}
\psfrag{N=8000}[r][r][0.9]{$8 \times 10^3$}
\psfrag{N=12000}[r][r][0.9]{$1.2\times 10^4$}
\psfrag{N=125000}[r][r][0.9]{$1.25\times 10^5$}
\psfrag{N=1000000}[r][r][0.9]{$10^6$}
\includegraphics[width=17cm,height=7.0cm]{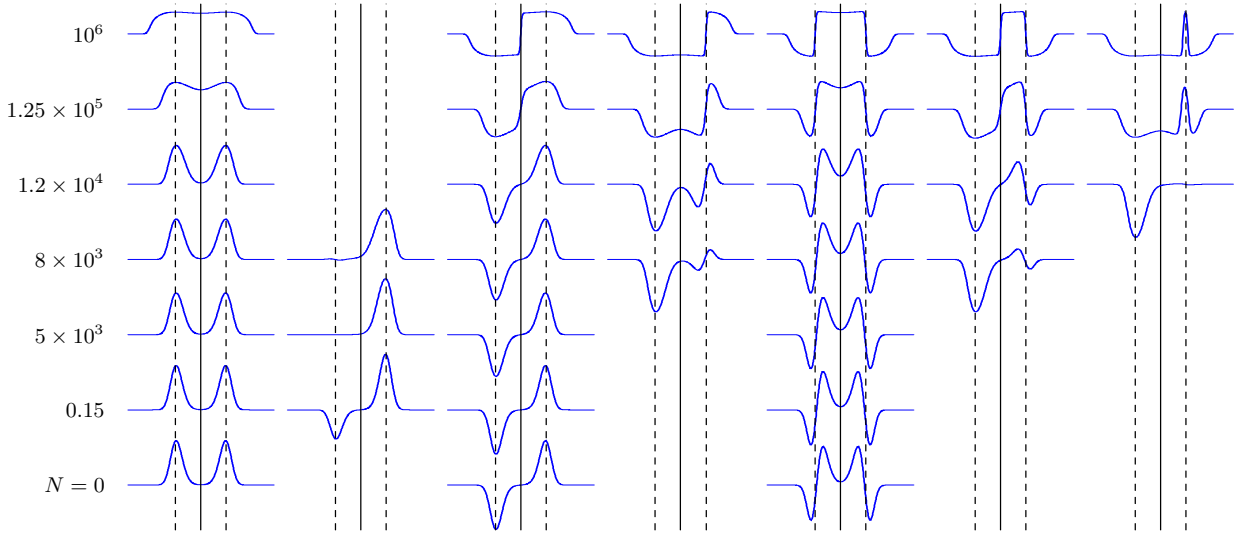}
\caption{Zero-, one-, and two-soliton stationary solutions of the 
repulsive GPE with the symmetric double-well potential 
(\protect{\ref{bistablep}}) 
for different values of the normalization $N$.
For comparison, the functions are shown scaled by $\sqrt{N}$.
The vertical solid and dashed lines indicate the double-well
maximum and minima, respectively.
The degenerate states obtained by changing $\psi(x) \to \psi(-x)$ 
are not reported.
The results have been obtained with the following parameters:
$m=3.818\times 10^{-26}~ {\rm Kg}$,
$\omega=12.75~ {\rm Hz}$,
$\gamma=10^9~ {\rm Kg^{-\frac{1}{4}}m^{-\frac{1}{2}}s^{-\frac{1}{2}}}$,
$U_0= 1.1087\times 10^{-41}~ {\rm Jm}$.}
\label{rgpestates}
\end{figure*}
\begin{figure*}   
\psfrag{N=0}[r][r][0.9]{$N=0$}
\psfrag{N=0.2}[r][r][0.9]{$0.2$}
\psfrag{N=50}[r][r][0.9]{$50$}
\psfrag{N=12000}[r][r][0.9]{$1.2\times 10^4$}
\psfrag{N=50000}[r][r][0.9]{$5\times 10^4$}
\psfrag{N=70000}[r][r][0.9]{$7\times 10^4$}
\includegraphics[width=17cm,height=7.0cm]{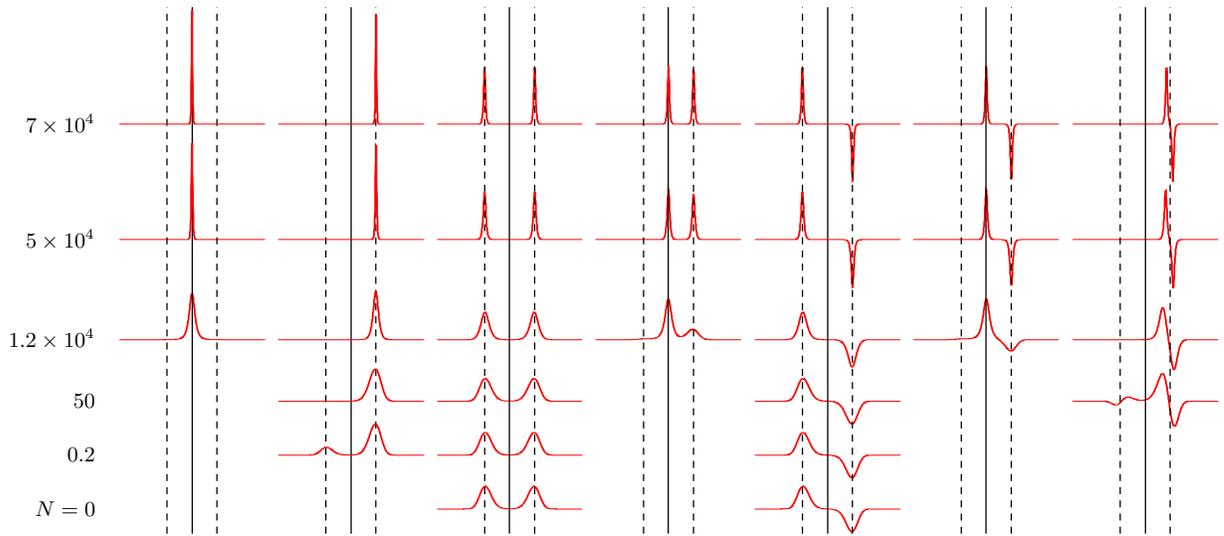}
\caption{As in Fig. \protect{\ref{rgpestates}} for the 
one- and two-soliton solutions in the attractive case
$U_0= -1.1087\times 10^{-41}~ {\rm Jm}$.}
\label{agpestates}
\end{figure*}

\section{Stationary solutions}
Let us first review some general properties of the stationary 
solutions of the GPE that reduce, in the limit of vanishing nonlinearity, 
to the eigenfunctions of the associated Schr\"odinger equation
\begin{equation}
\left[  - \frac{\hbar^2}{2m}\nabla^2 + V(\vb{x}) -{\cal E}_n \right]
\phi_n(\vb{x})=0,~n=0,1,\ldots.
\end{equation}
In \cite{dmp} we have shown that for any finite value of the chemical 
potential $\mu$ there exists a set of stationary solutions of the GPE,  
$\Psi_{\mu n}(\vb{x},t)=\exp \left( -\frac{i}{\hbar}\mu t \right) 
\psi_{\mu n}(\vb{x})$, 
which have limit 
$\psi_{\mu n}(\vb{x})~\|\psi_{\mu n}\|^{-1} \to \phi_n(\vb{x})$
when $\mu \to {\cal E}_n$.
The parameter $\mu$ ranges in the interval $[{\cal E}_n, +\infty)$  
for $U_0>0$ and in the interval
$(- \infty, {\cal E}_n]$ for $U_0<0$. 
In both cases, the number of particles in the state $\psi_{\mu n}$, 
$N_n(\mu)=\|\psi_{\mu n}\|^2$, vanishes for $\mu \to {\cal E}_n$.
In other words, the linear limit is reached for a vanishing number
of particles in the condensate.

In the 1-D case, asymptotically exact expressions for 
the GPE stationary solutions with linear counterpart are known
also in the opposite limit of strong nonlinearity.
For $\mu \to \pm \infty$, depending on the sign of $U_0$, 
these solutions assume the form of chains of dark or bright solitons
\cite{dmp}.
More specifically, in the repulsive case $U_0>0$ the solution with 
$n=0$ nodes assumes the zero-soliton shape
\begin{equation}
\psi_{\mu 0}(x) 
\to 
\left\{
\begin{array}{ll}
\sqrt{\left(\mu-V(x)\right)/U_0} & \mu>V(x) \\
0 & \mu<V(x)
\end{array} 
\right. ,
\label{tf0}
\end{equation}
while for $n\geq 1$ nodes we obtain asymptotic solutions
with  $n$ dark solitons
\begin{equation}
\psi_{\mu n}(x) \to \psi_{\mu 0}(x)\prod_{k=1}^{n}
\tanh \left(\frac{\sqrt{m\mu}}{\hbar}(x-x_k)\right).
\label{dsc}
\end{equation}
In the attractive case $U_0<0$, for $\mu \to - \infty$ the solutions 
with $n \geq 0$ nodes give rise to $n+1$ bright solitons 
\begin{equation}
\psi_{\mu n}(x) \to \sqrt{\frac{2 \mu}{U_0}}~
\sum_{k=0}^{n}(-1)^k 
\mbox{sech} \left(\frac{\sqrt{-2 m\mu}}{\hbar}(x-x_k)\right) .
\label{bsc}
\end{equation}
In the functions (\ref{dsc}-\ref{bsc}) with two or more solitons,
the solitons do not overlap, 
i.e. the distance between their centers $x_k$ is much larger than 
the dark-soliton width
$\hbar/\sqrt{m \mu}$ or the bright-soliton width
$\hbar/\sqrt{-2 m \mu}$ \cite{dmp}.
Note that any stationary solution is invariant under a global 
phase change and we do not consider this trivial degeneracy.

The stationary solutions of the GPE, for $\mu$ fixed, are 
the critical points of the grand-potential functional
\begin{eqnarray}
\Omega[\psi]&=&\int \biggl[\frac{\hbar^2}{2m}|\nabla\psi(\vb{x})|^2+
\frac{U_0}{2}|\psi(\vb{x})|^4 +\left(V(\vb{x})-\mu\right)
 \nonumber \\ && \times
|\psi(\vb{x})|^2 \biggr] \mathrm{d}\vb{x}.
\label{omega}
\end{eqnarray}
Since for $|\mu|$ large the GPE solutions with linear counterpart
assume the form (\ref{tf0}-\ref{bsc}) with some specified centers 
$\{x_k\}$, we look for more general multi-soliton solutions  
in which the soliton centers may assume different values.
The allowed $\{x_k\}$ can be determined by substituting the 
expressions (\ref{tf0}-\ref{bsc}) in (\ref{omega}) and
extremizing the resulting function $\Omega(\{x_k\})$.

\section{zero-, one-, and two-soliton solutions in a double-well}
\label{analytical}
Zero-soliton solutions exist only in the repulsive case $U_0>0$
and are given by Eq. (\ref{tf0}).
For $\mu$ sufficiently large, we have a node-less state which 
extends over the entire double well (column 1 of Fig. 1).
If $\mu$ is smaller than the barrier height 
$\omega^4/4 \gamma^4$, this state vanishes in the barrier region
where $V(x)> \mu$.
In this case, since $\psi=0$ is a trivial solution of the GPE, 
we could expect also two other stationary solutions of the form
$\psi(x)=\sqrt{\left(\mu-V(x)\right)/U_0}$ in one of the two 
wells and $\psi(x)=0$ elsewhere (column 2 of Fig. 1 and symmetric
partner $\psi(-x)$).
The new solutions break the symmetry of $V$ 
and must disappear in the linear limit.
They correspond to the self-trapped states studied in 
\cite{els,sfgs,mcww,oklhca}.

One-soliton solutions are described by Eq. (\ref{dsc}) with $n=1$ in the
repulsive case and Eq. (\ref{bsc}) with $n=0$ in the attractive one.
The corresponding grand-potential becomes a function of the soliton 
centers $x_1$ or $x_0$, respectively.
For $|\mu|$ sufficiently large, the width of the solitons
is very small and the dependence of the integral (\ref{omega}) 
on $x_1$ or $x_0$ is due only to the term $V|\psi|^2$.
The dark soliton density $|\psi_{\mu 1}|^2$ is constant with a hole 
in $x_1$ so that $\Omega(x_1) \sim {\rm const} -V(x_1)$.
The bright soliton density $|\psi_{\mu 0}|^2$ is different from zero
only in proximity of $x_0$ and $\Omega(x_0) \sim {\rm const} +V(x_0)$.
In both cases we have three one-soliton solutions corresponding to the 
three extrema of the external potential.
The soliton may be found in the maximum
(column 3 of Fig. 1 and column 1 of Fig. 2)
or in one of the two minima (column 4 of Fig. 1 and column 2 of Fig. 2
and symmetric partners $\psi(-x)$) of the double-well.
The two solutions with the soliton centers in $\pm x_m$,
where $x_m= \sqrt{\omega^2/2 m \gamma^4}$, 
break the symmetry of $V(x)$ and do not have linear counterpart.
 
In the repulsive case, two-soliton solutions are described by Eq. 
(\ref{dsc}) with $n=2$ and the grand-potential becomes the 
two-variable function $\Omega(x_1,x_2)$.
When the distance between the soliton centers is much larger 
than their width, we have
$\Omega(x_1,x_2) \simeq \Omega(x_1)+\Omega(x_2)$.
In the region $x_1 < x_2$, $\Omega$ has a maximum in 
$(-x_m,x_m)$ and two saddle points in $(0,x_m)$ and $(-x_m,0)$.
We assume that $x_m \gg \hbar/\sqrt{m \mu}$.
The stationary solution corresponding to the maximum of $\Omega$ 
is shown in column 5 of Fig. 1. 
Those corresponding to the two saddle points
(column 6 of Fig. 1 and symmetric partner $\psi(-x)$)
break the symmetry of $V$ and must disappear in the linear limit.

Other extrema of $\Omega$ can be found
when the centers of the two dark solitons are into the same well.
In fact, when both $x_1$ and $x_2$ tend to $x_m$, or $-x_m$,
the value of $\Omega(x_1,x_2) \sim {\rm const} -V(x_1)-V(x_2)$
increases until $|x_1-x_2| \gg\hbar/\sqrt{m\mu}$.
When the distance $|x_1-x_2|$ becomes comparable with the soliton width, 
the two density holes in $|\psi_{\mu 2}|^2$ begin to merge and
the norm of $\psi_{\mu 2}$ increases. 
This implies that $\Omega$ decreases for $|x_1-x_2| \to 0$
since, at least for $\mu$ sufficiently large, 
$\Omega \sim -\mu\| \psi_{\mu 2} \|^2$.
As a consequence, $\Omega$ has two maxima in
$(x_m-\delta,x_m+\delta)$ and $(-x_m-\delta,-x_m+\delta)$
with $2\delta \gtrsim \hbar/\sqrt{m\mu}$. 
The corresponding solutions 
(column 7 of Fig. 1 and symmetric partner $\psi(-x)$),
break the symmetry of $V(x)$ and do not have linear counterpart.

In the attractive case the situation is more complicated.
The bright solitons in the stationary solutions with linear 
counterpart given by Eq. (\ref{bsc}) are multiplied by a phase factor 
which is alternatively $+1$ and $-1$.
In general, we can expect bright solitons with arbitrary relative 
phases since each $\mbox{sech}$ function is, 
for $\mu \to - \infty$, solution of the GPE and 
this equation is invariant under a global phase change.
Restricting to real solutions, in the two-soliton case we have 
to consider the following possibilities
\begin{eqnarray}
\psi_{\mu 1}^\pm(x) &=& \sqrt{\frac{2 \mu}{U_0}}~
\left[
\mbox{sech} \left(\frac{\sqrt{-2 m\mu}}{\hbar}(x-x_0)\right)
\right. \nonumber \\ && \left.
\pm 
\mbox{sech} \left(\frac{\sqrt{-2 m\mu}}{\hbar}(x-x_1)\right)
\right].
\label{2bs}
\end{eqnarray}
The functions $\Omega^\pm(x_0,x_1)$ obtained by inserting these 
expressions in (\ref{omega}) present, in analogy with the repulsive
case, a minimum in $(-x_m,x_m)$ 
and two saddle points in $(0,x_m)$ and $(-x_m,0)$.
The stationary states corresponding to the minimum of 
$\Omega^\pm(x_0,x_1)$ are shown in columns 3 and 5 of Fig. 2.
Those corresponding to the two saddle points 
(columns 4 and 6 of Fig. 2 and symmetric partners $\psi^\pm(-x)$)
break the symmetry of $V$ and do not have linear counterpart.

On the other hand, due to the gradient term in (\ref{omega}),
we have a different behavior of $\Omega^+$ and $\Omega^-$ 
when both the soliton centers $x_0$ and $x_1$ 
move toward the minimum of the same well.
In fact, $\Omega^+$ does not present new extrema 
while $\Omega^-$ has two minima in 
$(x_m-\delta,x_m+\delta)$ and $(-x_m-\delta,-x_m+\delta)$
with $2\delta \gtrsim \hbar/\sqrt{-2 m \mu}$. 
The corresponding solutions 
(column 7 of Fig. 2 and symmetric partner $\psi^-(-x)$),
break the symmetry of $V(x)$ and do not have linear counterpart.
\begin{figure}   
\psfrag{N}[][][0.9]{$N$}
\psfrag{E/Nho}[][][0.9]{$E/N\hbar\omega$}
\includegraphics[width=7.5cm]{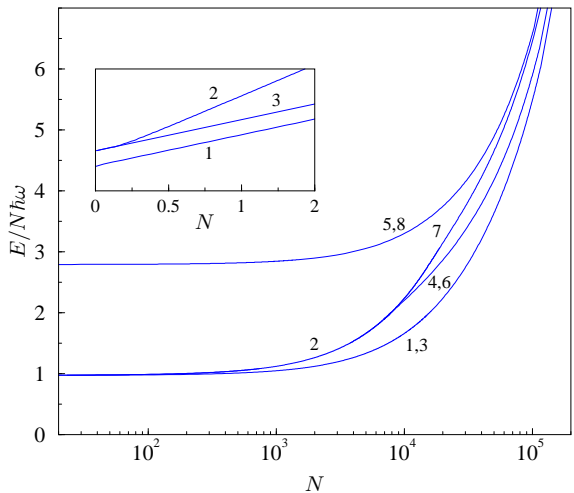}
\caption{Single-particles energies $E/N$ for the states
shown in Fig. \protect{\ref{rgpestates}} as a function of $N$.
The numbers correspond to the columns of 
Fig. \protect{\ref{rgpestates}}. The curve 8 corresponds to
the anti-symmetric partner of 5.}
\label{rgpeenergies}
\end{figure}
\begin{figure}
\psfrag{N}[][][0.9]{$N$}
\psfrag{E/Nho}[][][0.9]{$E/N\hbar\omega$}
\includegraphics[width=7.5cm]{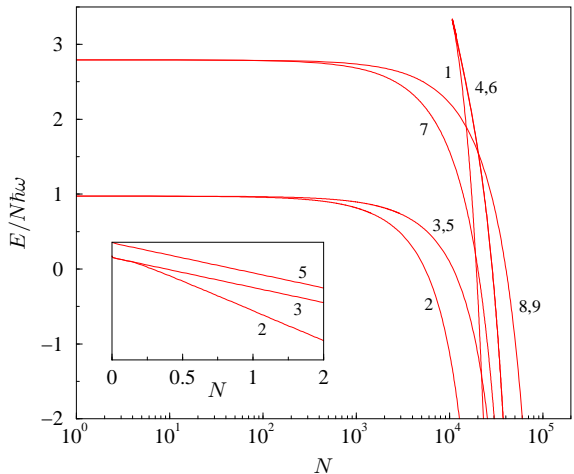}
\caption{As in Fig. \protect{\ref{rgpeenergies}} for the states
shown in Fig. \protect{\ref{agpestates}}.
Curves 8 and 9 correspond to the stationary solutions, 
not shown in Fig. \protect{\ref{agpestates}},
having as linear counterpart the Schr\"odinger eigenfunctions 
with 2 and 3 nodes, respectively.}
\label{agpeenergies}
\end{figure}

\section{Numerical solutions with an arbitrary number of particles}
\label{numerical}
Now we compare the zero-, one-, and two-soliton solutions
discussed above with the results of numerical simulations.
We use a numerical algorithm based on a standard relaxation method 
for partial differential equations \cite{numrec}.
The success of this method is crucially based on the quality
of the trial functions used to start the relaxation.
For $|\mu|$ very large, good trial functions are represented by the
multi-soliton functions with the soliton centers
determined as above. 
The relaxed solutions can be then used as 
trial functions for a new simulation with a smaller value of $|\mu|$.
By changing $\mu$ sufficiently slowly, one can follow
the evolution of the stationary states until they reach the linear 
limit, if it exists, or the point where they disappear.
Figures \ref{rgpestates} and \ref{agpestates} show in the repulsive 
and attractive cases, respectively, the states obtained in this way 
for different values of their norm $N$.
For any node index $n$, the linear limit is reached
when $N=N_n(\mu) \to 0$. 
All the solutions that break the symmetry of the external potential
disappear for $N$ smaller than a critical value. 
However, there also exist solutions without linear counterpart
that preserve this symmetry.
An example is shown in the first column of Fig. 
\ref{agpestates} which corresponds to a bright soliton at the 
center of the barrier.

In Figs. \ref{rgpeenergies} and \ref{agpeenergies} we show the 
single particle energies for the same states of Figs.
\ref{rgpestates} and \ref{agpestates} as a function of $N$.
From these figures it is evident the generation of
solutions without linear counterpart as $N$ is increased. 
In the case of the attractive GPE, the stationary solution which
for $N$ large is fully localized into one of the two wells (second 
column of Fig. \ref{agpestates}) is, when it exists, 
the state of minimal energy. 
Therefore, the nature of the mean-field ground state changes
as a function of $N$ and this suggests the existence of a quantum 
phase transition in the corresponding exact many-body system.   

The generation of stationary states without linear counterpart
can be understood in terms of bifurcations of superpositions of 
Schr\"odinger eigenstates. 
In the following we discuss an analytical example valid when 
the zero point energy of each isolated well, 
$\frac{1}{2} \hbar 2 \omega$, 
is much smaller than the barrier height, 
$\omega^4/4 \gamma^4$, i.e. for $\omega^3/\hbar\gamma^4\gg 1$. 
Let us consider  stationary solutions of the GPE of the form
\begin{equation}
\psi(x)=\sqrt{N}\left[a_0\chi_0(x-x_m)+b_0\chi_0(x+x_m)\right],
\label{abb}
\end{equation}
where $\chi_n(x)$ are the eigenfunctions of the Schr\"odinger problem 
with harmonic potential $\frac12 m (2\omega)^2 x^2$
and $a_0^2+b_0^2=1$.
Since the state (\ref{abb}) is normalized to $N$, for it to be a
stationary solution of the GPE we have to extremize 
the energy functional $E[\psi]=\Omega[\psi]+ \mu N$.
Up to exponentially small terms we get
\begin{equation}
E(b_0) \sim b_0\sqrt{1-b_0^2} +
\mbox{sign}(U_0) \frac{N}{N_0} \left( 1+2 b_0^4-2b_0^2 \right),
\end{equation}
where
\begin{equation}
N_0\sim \frac{\omega^3}{\hbar \gamma^4}
\exp\left( {-\frac{\omega^3}{\hbar \gamma^4}} \right) 
\sqrt{\hbar^3\omega/mU_0^2}.
\label{N0}
\end{equation}
For $N \ll N_0$, $E(b_0)$ has a minimum for $b_0=2^{-\frac12}$ and a 
maximum for $b_0=-2^{-\frac12}$. 
These extrema correspond to the lowest energy 
symmetric and anti-symmetric linear states
(columns 1 and 3 of Fig. \ref{rgpestates} 
and columns 3 and 5 of Fig. \ref{agpestates}).
If $U_0>0$, for $N \simeq N_0$ the maximum at $b_0=-2^{-\frac12}$ 
bifurcates in a minimum and a maximum which, increasing $N$, 
moves to $b_0=0$.
This describes the birth of the state in the second column of 
Fig. \ref{rgpestates} and its subsequent localization in the right 
well ($a_0=1$).  
If $U_0<0$ a similar result is obtained with maxima and minima exchanged
(see column 2 of Fig. \ref{agpestates}).
Generation of other states can be 
obtained by considering superpositions more complicated than (\ref{abb}).

\section{Stability of stationary solutions}\label{stability}

In this Section we discuss the stability of the stationary 
states described above.
We start with a linear stability analysis.
Consider the linearization of Eq. (\ref{gpe}) 
for a small change $\delta\Psi$ of its solution
\begin{equation}
i\hbar\frac{\partial}{\partial t} \delta\Psi =
\left[-\frac{\hbar^2}{2m}\nabla^2+V(\vb{x})+2U_0|\Psi|^2\right]
\delta\Psi + U_0\Psi^2\delta\Psi^*.
\label{gpev}
\end{equation}
We are interested to evaluate the evolution of the variation $\delta\Psi$
of a stationary solution $\Psi_{\mu n}$.
By writing
\begin{eqnarray}
\Psi+\delta\Psi &=&
\Psi_{\mu n}(\vb{x},t) + \delta\Psi(\vb{x},t) \nonumber \\&=& 
e^{-\frac{i}{\hbar}\mu t} 
\left[ \psi_{\mu n}(\vb{x}) + \delta\phi(\vb{x},t) \right],
\label{psiv}
\end{eqnarray}
according to Eq. (\ref{gpev}) the variation $\delta\phi$ and its 
complex conjugated $\delta\phi^*$ are determined by 
\begin{equation}
i\hbar\frac{\partial}{\partial t} 
\left( \begin{array}{l} 
\delta\phi \\
\delta\phi^*
\end{array}\right)
=
\left(\begin{array}{cc}
\mathcal{D}_{\mu n} & U_0{\psi_{\mu n}}^2 \\
-U_0{\psi_{\mu n}^*}^2 & -\mathcal{D}_{\mu n}
\end{array}\right)
\left( \begin{array}{l} 
\delta\phi \\
\delta\phi^*
\end{array}\right),
\label{lse}
\end{equation}
where
\begin{equation}
\mathcal{D}_{\mu n} = 
-\frac{\hbar^2}{2m}\nabla^2+V(\vb{x})+2U_0|\psi_{\mu n}|^2-\mu.
\end{equation}
The solution of Eq. (\ref{lse}) can be written as
\begin{equation}
\left( \begin{array}{l} 
\delta\phi (\vb{x},t)\\
\delta\phi(\vb{x},t)^*
\end{array}\right)
= \sum_k c_k e^{-i\lambda_k t /\hbar}
\left( \begin{array}{l} 
f_k(\vb{x}) \\ g_k(\vb{x})
\end{array}\right),
\label{tori}
\end{equation}
where $\lambda_k$ and $(f_k,g_k)$ are the eigenvalues and the 
eigenvectors of the linearization operator
\begin{equation}
\left(\begin{array}{cc}
\mathcal{D}_{\mu n} & U_0{\psi_{\mu n}}^2 \\
-U_0{\psi_{\mu n}^*}^2 & -\mathcal{D}_{\mu n}
\end{array}\right)
\left( \begin{array}{l} 
f_k \\ g_k
\end{array} \right)
= \lambda_k 
\left( \begin{array}{l} 
f_k \\ g_k
\end{array}\right)
\label{spectrum}
\end{equation}
and the coefficients $c_k$ are fixed by the initial condition
$\delta\phi(\vb{x},0)$.
Multiplying (\ref{spectrum}) by $\left(f_k^*,-g_k^*\right)$ and 
integrating over space, we get
\begin{eqnarray}
&&\int \left[ f_k^* \mathcal{D}_{\mu n} f_k 
+ g_k^* \mathcal{D}_{\mu n} g_k
+ U_0 \left(f_k^* g_k + f_k g_k^* \right) \right] \mathrm{d}\vb{x}
\nonumber\\&&\qquad=
\lambda_k \int \left(|f_k|^2 - |g_k|^2 \right)\mathrm{d}\vb{x}.
\end{eqnarray}
The linearization eigenvalues are real and 
Eq. (\ref{lse}) admits quasiperiodic solutions 
for an arbitrary $\psi_{\mu n}$ \cite{nota}.  
This proves the linear stability of the stationary solutions.

Lyapunov stability of all the states $\Psi$ in the neighborhood of
a stationary solution $\Psi_{\mu n}$ is a mathematically stronger 
concept of stability and certainly more relevant from an
experimental point of view.  
This kind of stability was previously studied in the case of a 
lattice model which reduces to the GPE in the continuum limit \cite{cjp}.
In that paper it was shown numerically that the maximum Lyapunov exponent 
associated to a discretized version of Eq. (\ref{gpev}) vanishes
when the initial state $\Psi(\vb{x},0)$ is sufficiently close 
to one of the stationary states.
A similar analysis can be pursued in the case of the GPE by simulating
a huge finite dimensional system, namely that obtained by applying a  
finite difference scheme to the partial differential equation (\ref{gpe}).
Of course, the large but finite number of degrees of freedom used
in the simulation sets a limit to the maximum time at which the 
properties of the infinite dimensional system corresponding to the 
GPE are correctly represented.  
We will report on this elsewhere.
Here we note that the gained scenario is consistent with Kuksin theory 
\cite{kuksin} which asserts that 
Eq. (\ref{gpev}) admits $N$-dimensional invariant tori, 
deformation of (\ref{tori}), 
in a finite neighborhood of any stationary state
whose linearization spectrum satisfy non-degeneracy and 
non-resonance conditions.

The spectrum of the linearization operator is useful also for 
discussing the stability of the stationary states under the
effect of a dissipative perturbation.
The grand-potential (\ref{omega}) evaluated for a state of the form
(\ref{psiv}), up to the second order in the variation $\delta\phi$
gives
\begin{equation}
\Omega[\Psi+\delta\Psi] = \Omega[\psi_{\mu n}]+ \delta^2\Omega,
\end{equation}
where
\begin{eqnarray}
\delta^2 \Omega &=&
\frac12
\int \delta\phi^* \left[ \mathcal{D}_{\mu n} \delta\phi
+U_0{\psi_{\mu n}}^2 \delta\phi^* \right] \mathrm{d}\vb{x}
\nonumber\\&&+
\frac12
\int \delta\phi \left[ \mathcal{D}_{\mu n} \delta\phi^*
+U_0{\psi_{\mu n}^*}^2 \delta\phi \right] \mathrm{d}\vb{x}.
\label{d2omega}
\end{eqnarray}
By using Eqs. (\ref{lse}) and (\ref{tori}) and the sum rule
\begin{equation}
\sum_k c_ke^{-i\lambda_k t/\hbar}f_k=\sum_k
c_k^*e^{i\lambda_k t/\hbar}g_k^*,
\end{equation}  
we get 
\begin{equation}
\delta^2 \Omega = \frac{1}{2} \sum_k |c_k|^2 \lambda_k
\left(\|f_k\|^2 - \|g_k\|^2 \right). 
\end{equation}
Therefore, a stationary solution $\psi_{\mu n}$ is a local minimum 
of the grand-potential functional if and only if for any $k$
we have \cite{garcia}
\begin{equation}
\lambda_k \left(\|f_k\|^2 - \|g_k\|^2 \right) \geq 0.
\label{esc}
\end{equation} 

To verify the disequalities (\ref{esc}),
we have solved numerically the eigenvalue problem (\ref{spectrum})
by representing the linearization operator with a finite difference 
scheme. 
In the repulsive case $U_0>0$, the condition (\ref{esc}) is fulfilled 
only by the state in the first column of Fig. 1.
In the attractive case $U_0<0$, no one of the states shown in Fig. 2
satisfies (\ref{esc}). 
This can be explained observing that the grand-potential evaluated
at a stationary state is
\begin{equation}
\Omega[\psi_{\mu n}] = - \frac{1}{2}~U_0 
\int \left| \psi_{\mu n} \right|^4 \mathrm{d}\vb{x}.
\end{equation}
Thus, if $U_0<0$, $\Omega$ assumes the minimal value for 
the trivial solution $\psi=0$.
These results have a certain interest on the stability 
of a physical condensate in which a dissipative dynamic is introduced
by the coupling with the environment degrees of freedom.
Eventually the system will converge to a local minimum 
of $\Omega$.
For attractive interaction, this implies the disappearance of 
the condensate.
An estimate of the characteristic lifetimes has been given 
in the case of a vortex state \cite{fs}. 
 
We conclude our stability analysis by considering the short time behavior
of the stationary states under the action of an initial finite 
deformation. 
A similar analysis has been considered in \cite{ckr} to check the
stability of the solutions found in \cite{ccr}.
The authors of \cite{ckr} studied the evolution of stationary states
initially perturbed with a stochastic noise.
In our case, the stationary solutions corresponding to solitons located
near the extrema of the double-well potential should have great 
sensitivity, specially in the case of unstable extrema,
to symmetry breaking perturbations.
Here, we consider the evolution of shifted stationary states,
i.e. we solve Eq. (\ref{gpe}) with the initial condition
$\Psi(x,0)=\psi_{\mu n}(x-\Delta x)$.
The numerical simulations have been performed with the 
improved Crank-Nicholson scheme introduced in \cite{cjp} 
which provides an accurate conservation of the constants of motion 
of (\ref{gpe}), namely norm and energy.  
As an example, we describe the evolution of the states 
of column 4 of Fig. 1 and 2.
Note that these are states without linear counterpart. 
\begin{figure}   
\psfrag{t (s)}[][][0.9]{$t~(s)$}
\psfrag{x1/xm}[][][0.9]{$x_1/x_m$}
\psfrag{x0/xm, x1/xm}[][][0.9]{$x_0/x_m$, $x_1/x_m$}
\includegraphics[width=7.5cm]{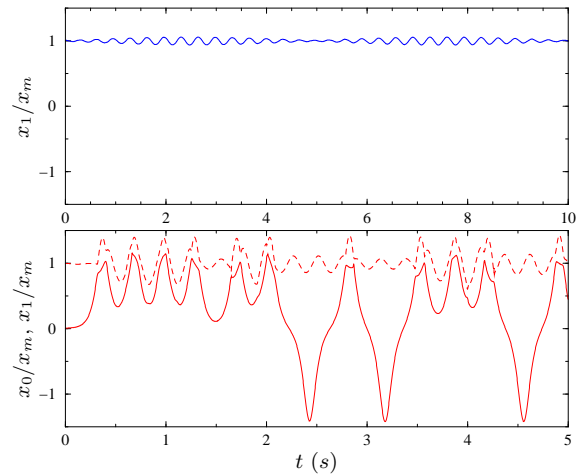}
\caption{Time evolution of the soliton centers for initially 
perturbed stationary states $\Psi(x,0)=\psi_{\mu n}(x-\Delta x)$ 
with $\Delta x=0.3$ $\mu$m.
In the upper panel $\psi_{\mu n}$ is one of the states 
in column 4 of Fig. 1 while in the lower panel one of those 
in column 4 of Fig. 2.}
\label{shift}
\end{figure}

The state with one dark soliton has a rather simple evolution. 
The qualitative shape of the state does not change but 
the soliton oscillates around the minimum $x=x_m$ of the right well.
The position of the soliton center, $x_1$, as a function of time is shown
in the upper panel of Fig. \ref{shift} for $\Delta x=0.3$ $\mu$m.
The amplitude of the oscillations is very small in the case considered
and increases by increasing $\Delta x$.  
For a shift $\Delta x$ sufficiently large the soliton can jump 
between the two wells.

The dynamics of the state with two bright solitons is more complicated,
see lower panel of Fig. \ref{shift}.
Initially the soliton at the center of the barrier moves toward 
that located inside the right well, the latter being essentially 
at rest.  
When the distance between the two solitons becomes comparable 
to their width the oscillations of the solitons inside the same
well turn out to be correlated. 
The solitons do not cross each other.
When the potential energy of the barrier, $V|\Psi|^2$, is sufficiently 
reduced by the negative interaction energy, $\frac12 U_0 |\Psi|^4$,
the soliton which was originally at the center of the barrier 
can jump into the left well.
This inter-well dynamics is obtained also for values of the initial
shift smaller than $\Delta x=0.3$ $\mu$m which is the case shown 
in Fig. \ref{shift}.
By decreasing $\Delta x$, the initial falling of the soliton 
at the center of the barrier into the right well 
(left well for $\Delta x<0$) is slowed down.
 
The results shown in Fig. \ref{shift} can be generalized to
different kinds of perturbations, e.g. stochastic noise, modification
of the parameters of the external potential.
Details will be reported elsewhere.

\section{Conclusions}
We have shown that in presence of an external potential a 1-D GPE
can admit stationary solutions without linear counterpart.
Their existence is strictly connected to the multi-well nature 
of the potential.
In the double well example discussed here, these solutions disappear
in the limit $\omega \to 0$ when the potential assumes the shape 
of a single quartic well.
For a piece-wise constant double-well, the stationary states here 
discussed analytically only in the limit of strong nonlinearity can be 
obtained in terms of Jacobi elliptic functions for any number of 
particles in the condensate. 

We have also discussed the stability of the stationary states
under different points of view. 
The results indicate that the soliton-like states, with and without 
linear counterpart, are sufficiently stable on the typical time scales
of a BEC experiment.
By voluntarily introducing perturbations of proper intensity,
a soliton dynamics could also be observed.

\begin{acknowledgments}
We thank R. Onofrio for very useful comments on the experimental aspects 
of BEC and a critical reading of the manuscript.
This work was supported in part by Cofinanziamento MURST 
protocollo MM02263577\_001.
\end{acknowledgments}


\begin{thebibliography}{99}
\bibitem{nota} A linearization eigenvalue $\lambda_{\bar{k}}$ for which
$\|f_{\bar{k}}\|=\|g_{\bar{k}}\|$ can be complex. 
This restricts quasiperiodic evolution to initial variations
$\delta\phi(\vb{x},0)$ for which $c_{\bar{k}}=0$.
\end{thebibliography}
\end{document}